\documentclass[preprint,aps,prd,floatfix,nofootinbib,11pt]{revtex4-2}
\pdfoutput=1
\usepackage{graphicx}
\usepackage{amsfonts}
\usepackage{amssymb}
\usepackage{xcolor}
\usepackage{natbib}
\usepackage{braket}

\usepackage{hyperref}

\usepackage{amsmath}
\usepackage{rotating}
\usepackage{amssymb,amsthm}
\usepackage{soul}

\usepackage{xcolor}

\usepackage{latexsym}
\usepackage{graphics}

\usepackage{amsfonts}
\usepackage{amsmath}
\usepackage{rotating}
\usepackage{amssymb}

\usepackage{amsmath}
\usepackage{rotating}
\usepackage{amssymb}
\usepackage{soul}

\usepackage{xcolor}
\usepackage{xcolor}
\def\beq{\begin{eqnarray}}  
	\def\eeq{\end{eqnarray}}

\def \bh {\mbox{{\bf h}}}

\begin{document}
	
	\title{Scalar field source Teleparallel Robertson-Walker $F(T)$-gravity solutions}

	\author{A. Landry}
	\email{a.landry@dal.ca}
	\affiliation{Department of Mathematics and Statistics, Dalhousie University, Halifax, Nova Scotia, Canada, B3H 3J5}

%add authors here

\begin{abstract}
This paper investigates the teleparallel Robertson--Walker (TRW) $F(T)$ gravity solutions for a scalar field source. We use the TRW $F(T)$ gravity field equations (FEs) for each $k$-parameter value case added by a scalar field to find new teleparallel $F(T)$ solutions. For $k=0$, we find an easy-to-compute $F(T)$ solution formula applicable for any scalar field source. Then, we obtain, for $k=-1$ and $+1$ situations, some new analytical $F(T)$ solutions, only for specific $n$-parameter values and well-determined scalar field cases. We can find by those computations a large number of analytical teleparallel $F(T)$ solutions independent of any scalar potential $V(\phi)$ expression. The $V(\phi)$ independence makes the FE solving and computations easier. The new solutions will be 	 relevant for future cosmological applications in dark matter, dark energy (DE) quintessence, phantom energy and quintom models of physical processes.
\end{abstract}

\maketitle

%\newpage

%\tableofcontents

\newpage

\section{Introduction}\label{sect1}

The teleparallel theories of gravity are an important and promising class of alternative theories where all quantities and symmetries are defined in terms of the coframe and spin-connection \cite{Krssak:2018ywd,Krssak_Saridakis2015,Coley:2019zld,HJKP2018,HJKP2018a,Bahamonde:2021gfp,Cai_2015,preprint,MCH,SSpaper,FTBcosmogholamilandry,TdSpaper,coleylandrygholami,nonvacSSpaper,nonvacKSpaper,roberthudsonSSpaper,scalarfieldKS,Aldrovandi_Pereira2013}. The most appropriate definition of teleparallel geometry is affine symmetry. Frame-based symmetry on the frame bundle is defined by a coframe/spin-connection pair and a field ${\bf X}$ satisfying some fundamental Lie-derivative relations \cite{Coley:2019zld,MCH,preprint,coleylandrygholami}. Such relationship definitions are considered the {\it {frame-dependent}} %MDPI: Please confirm if the italics is unnecessary and can be removed. Please check all italics format in the MS. AL: OK.
analogues of the definition of symmetry, as presented in refs. \cite{HJKP2018,HJKP2018a}. For a pure teleparallel geometry, a coframe/spin-connection pair will also have satisfy the null Riemann curvature criteria relation, $R^a_{~b\mu\nu}=0$, which leads to the spin-connection solution $\omega^a_{~b\mu} = \Lambda^a_{~c}\partial_{\mu}\Lambda_b^{~c}$ in terms of a typical Lorentz transformation $\Lambda^a_{~b}$ \cite{Coley:2019zld}.

By using the Cartan--Karlhede (CK) algorithm, we can construct some invariant coframe/spin-connection pairs satisfying affine frame symmetries. However, there are cosmological teleparallel spacetime geometry classes that are invariant under the full $G_6$ Lie algebra of affine symmetries \cite{preprint,coleylandrygholami,FTBcosmogholamilandry}. In addition, the proper coframe and the pure symmetric field equations (FEs) have always been %Please check the intended meaning is retained. AL: OK.
obtained and described by a teleparallel Robertson--Walker (TRW) geometry, the Robertson--Walker (RW) metric $g_{\mu\nu}$ and the parameter $k = (-1,0,1)$~{\cite{preprint,coleylandrygholami,FTBcosmogholamilandry}.} The parameter $k$ is defined as the constant spatial curvature in the RW pseudo-Riemannian metric, but $R^a_{~b\mu\nu}=0$ in teleparallel spacetimes. The parameter $k$ is usually considered as a three-dimensional space curvature in the metric-based approach and as a part of the torsion scalar in fourth-dimensional teleparallel spacetimes. In TRW geometries, an appropriate coframe--spin-connection pair leads to trivial antisymmetric FE parts. There are some TRW geometries where the coframe--spin-connection pair admits the full $G_6$ Lie algebra defined by the $6$ Killing Vectors (KVs). In the literature, there are papers for $k=\pm 1$ cases, with some investigated geometries not yielding a $G_6$ Lie algebra. There are also some solutions involving the use of inappropriate coframes and/or spin-connections.%Please check the intended meaning is retained
Recently, new coframe--spin-connection pairs satisfying a $G_6$ symmetry group have been constructed~{\cite{Hohmann:2015pva,HJKP2018,HJKP2018a,Hohmann:2018rwf}.} The most recent achievements in this regard, leading to new teleparallel solutions, are presented in refs.%Please check the intended meaning is retained
\cite{Coley:2019zld,preprint,coleylandrygholami,FTBcosmogholamilandry}.

A lot of works in the literature have been based on $k=0$ TRW cosmological model cases (refs. \cite{Bahamonde:2021gfp,Cai_2015} and references within). In particular, specific forms for $F(T)$ have been investigated by using a specific ansatz, and reconstruction methods have been explored extensively (where the function $F(T)$ is reconstructed from assumptions on the models). Dynamical systems methods, especially fixed point and stability analysis, in flat TRW models have been used, with the stability conditions studied via the standard de Sitter fixed point  \cite{Bahamonde:2021gfp,coley03,BahamondeBohmer,Kofinas,BohmerJensko,aldrovandi2003}. Then, $k\neq 0$ solutions have been recently studied in bounce and inflation models \cite{bounce,Capozz} (i.e., the analysis is only applicable for the case where $k=1$). The perturbations have also been studied in non-flat cosmology \cite{inflat}. In a recent paper, we found, as exact $k=0$ solutions for the FEs, a combination of two power-law terms with the cosmological constant for a linear perfect fluid \cite{coleylandrygholami}. For $k=\pm 1$ cases, the differential equation has been linearized and a rigorous stability test has been performed for determining specific conditions on possible non-flat cosmological solutions. This test has been conducted with the $k=0$ exact solution (dominating term) and a linear correction term by using a power-law ansatz, a perfect fluid Equation of State (EoS) and cosmological parameters for a stable TRW solution and model requirements \cite{coleylandrygholami}. After the development of $F(T)$ TRW geometries and solutions for linear perfect fluids, we can go further by using the same approach. There are new papers on cosmological teleparallel $F(T,B)$ solutions for linear and non-linear perfect fluids and also for some scalar field sources \cite{FTBcosmogholamilandry}. We can also add recent papers on Kantowski--Sachs (KS) (pure time-dependent spacetimes) teleparallel $F(T)$ solutions for perfect fluids and scalar fields \cite{nonvacKSpaper,roberthudsonSSpaper,scalarfieldKS}. Therefore, as we have already suggested, and as with the achievable research work presented in ref. \cite{coleylandrygholami}, studying scalar field TRW $F(T)$ solutions deserves to be the primary aim of the current paper. But there are also physical motivations for this new~paper. %Please check the intended meaning is retained. AL: OK.

The most important physical motivations concern the possible teleparallel dark energy (DE) models. The first motivation is the quintessence DE physical process described by the linear perfect fluid EoS $P_{\phi}=\alpha_Q\,\rho_{\phi}$, where $-1<\alpha_Q<-\frac{1}{3}$ \cite{steinhardt1,steinhardt2,steinhardt3,carroll1,quintessencecmbpeak,quintessenceholo,quintchakra2024,quintessencephantom,cosmofate,steinhardt2024}. This perfect-fluid-described DE model assumes a fundamental scalar field induces and explains this physical process and constitutes the first possible form of DE that satisfies the $P+\rho \geq 0$ energy condition \cite{scalarfieldKS}. There are also some teleparallel extension theories based on the scalar field as a boundary variable or scalar--torsion theories, to name only a few \cite{ftphicosmo,scalar2}. As a lower limit, there is the cosmological constant defined by a $\alpha_Q=-1$ perfect fluid as another form of DE; this constitutes some boundary of the quintessence process in terms of EoS. The other main DE form is the phantom energy (or negative energy), where $\alpha_Q<-1$, and the energy condition will usually be violated ($P+\rho \ngeq 0$) \cite{strongnegative,caldwell1,farnes,baumframpton,phantomteleparallel1,ripphantomteleparallel2,phantomteleparallel3,phantomdivide}. A phantom energy-based cosmological model is described as a strong, accelerated, expanding universe, leading---after a finite amount of time---to the Big Rip physical process (great breakdown) and constituting an extreme cosmological scenario. This often occurs when we assume a non-linear perfect fluid teleparallel cosmological model \cite{nonvacKSpaper,scalarfieldKS}. After determining the type of DE and study, it is interesting to observe the combination (or mix) of the quintessence and phantom DE models with the cosmological constant as an intermediate limit: the quintom physical process models~{\cite{quintom1,quintom2,quintom3,quintom4,quintomcoleytot,quintomteleparallel1}.} This is often described by two scalar field models: one field for quintessence and the second for phantom, or one for the unified process and a second for the coupling. There is even a study of a quintom oscillating model between quintessence and phantom energy, where the cosmological constant state is the mid-point oscillating amplitude position \cite{quintom3}. However, there is only one relevant paper concerning teleparallel quintom models \cite{quintomteleparallel1}. All these DE physical process works provide ample justification for studying possible scalar field TRW $F(T)$ solution classes.

For this paper, we will first summarize, in Section \ref{sect2}, the teleparallel $F(T)$ gravity FEs, the TRW coframe--spin-connection pair used and the scalar field source parameter equations. In Section \ref{sect3}, we will solve the FEs for flat cosmological cases ($k=0$), allowing some easy-to-compute $F(T)$ solutions and {\color{black} plotting most of the new solutions}. We will follow, in Sections \ref{sect4} and \ref{sect42}, with $k=-1$ and $+1$ solutions, but the FEs will only allow analytical $F(T)$ solutions for some specific subcases. We will conclude this new development in Section \ref{sect5}, addressing the possible impacts in favor of future DE models like quintessence, phantom and quintom physical processes.

\section{Summary of Teleparallel Gravity and Field Equations}\label{sect2}

\subsection{Teleparallel $F(T)$ Gravity Theory}

The teleparallel $F(T)$-type gravity action integral with any gravitational source {is as follows \cite{Aldrovandi_Pereira2013,Bahamonde:2021gfp,Krssak:2018ywd,SSpaper,nonvacSSpaper,nonvacKSpaper,roberthudsonSSpaper}:}
\begin{equation}\label{1000}
	S_{F(T)} = \int\,d^4\,x\,\left[\frac{h}{2\kappa}\,F(T)+\mathcal{L}_{Source}\right],
\end{equation}
where $h$ is the coframe determinant, $\kappa$ is the coupling constant and $\mathcal{L}_{Source}$ is the gravitational source term. We will apply the least-action principle to Equation \eqref{1000} to find the symmetric and antisymmetric parts of FEs as follows \cite{SSpaper,nonvacSSpaper,nonvacKSpaper,roberthudsonSSpaper}:
%\begin{subequations}
\begin{eqnarray}
	\kappa\,\Theta_{\left(ab\right)} &=& F_T\left(T\right) \overset{\ \circ}{G}_{ab}+F_{TT}\left(T\right)\,S_{\left(ab\right)}^{\;\;\;\mu}\,\partial_{\mu} T+\frac{g_{ab}}{2}\,\left[F\left(T\right)-T\,F_T\left(T\right)\right],  \label{1001a}
	\\
	0 &=& F_{TT}\left(T\right)\,S_{\left[ab\right]}^{\;\;\;\mu}\,\partial_{\mu} T, \label{1001b}
\end{eqnarray}
%\end{subequations}
with $\overset{\ \circ}{G}_{ab}$ being the Einstein tensor, $\Theta_{\left(ab\right)}$ the energy-momentum, $T$ the torsion scalar, $g_{ab}$ the gauge metric, $S_{ab}^{\;\;\;\mu}$ the superpotential (torsion-dependent) and $\kappa$ the coupling constant. The canonical energy momentum and its GR conservation law are obtained from the $\mathcal{L}_{Source}$ term of Equation \eqref{1000} as follows \cite{Aldrovandi_Pereira2013,Bahamonde:2021gfp}:
%\begin{subequations}
\begin{align}
	\Theta_a^{\;\;\mu}=&\frac{1}{h} \frac{\delta \mathcal{L}_{Source}}{\delta h^a_{\;\;\mu}}, \label{1001ca}
	\\
	\Rightarrow\quad &\overset{\ \circ}{\nabla}_{\nu}\left(\Theta^{\mu\nu}\right)=0 , \label{1001e}
\end{align}
%\end{subequations}
where $\overset{\ \circ}{\nabla}_{\nu}$ the covariant derivative and $\Theta^{\mu\nu}$ is the conserved energy momentum tensor. Equation \eqref{1001ca}'s antisymmetric and symmetric parts are as follows \cite{SSpaper}:
\begin{equation}\label{1001c}
	\Theta_{[ab]}=0,\qquad \Theta_{(ab)}= T_{ab},
\end{equation}
where $T_{ab}$ is the symmetric part of $\Theta^{\mu\nu}$. Equation \eqref{1001e} also imposes the symmetry of $\Theta^{\mu\nu}$ and then Equation \eqref{1001c}'s condition. Equation \eqref{1001c} is only valid when the matter field interacts with the metric $g_{\mu\nu}$ defined from the coframe $h^a_{\;\;\mu}$ and the gauge $g_{ab}$, and is not directly coupled to the $F(T)$ gravity. This consideration is only valid for the null hypermomentum case (i.e., $\mathfrak{T}^{\mu\nu}=0$), as discussed in refs. \cite{golov3,nonvacSSpaper,nonvacKSpaper,roberthudsonSSpaper}. This last condition on hypermomentum is defined in Equations \eqref{1001a}--\eqref{1001b} as follows \cite{golov3}:
\begin{align}\label{1001h}
	\mathfrak{T}_{ab}=\kappa\Theta_{ab}-F_T\left(T\right) \overset{\ \circ}{G}_{ab}-F_{TT}\left(T\right)\,S_{ab}^{\;\;\;\mu}\,\partial_{\mu} T-\frac{g_{ab}}{2}\,\left[F\left(T\right)-T\,F_T\left(T\right)\right]=0.
\end{align}
There are more general teleparallel $\mathfrak{T}^{\mu\nu}$ definitions and $\mathfrak{T}^{\mu\nu}\neq 0$ conservation laws, but this does not really concern the teleparallel $F(T)$ gravity situation \cite{hypermomentum1,hypermomentum2,hypermomentum3,golov3}.

\subsection{Teleparallel Robertson--Walker Spacetime Geometry}

In teleparallel gravity, any frame-based geometry on a frame bundle defined by a coframe--spin-connection pair and a {field} ${\bf X}$ must satisfy the fundamental Lie derivative-based equations \cite{Coley:2019zld,MCH,preprint,coleylandrygholami}:
\begin{equation}
	\mathcal{L}_{{\bf X}} \bh_a = \lambda_a^{~b} \,\bh_b { and } \mathcal{L}_{{\bf X}} \omega^a_{~bc} = 0, \label{Intro:FS2}
\end{equation}
where $\omega^a_{~bc}$ is the spin connection in terms of the differential coframe $\bh_a$ and $\lambda_a^{~b}$ is the linear isotropy group component. For TRW spacetime geometries on an orthonormal frame, the coframe $h^a_{\;\;\mu}$ solution will be as follows \cite{preprint,coleylandrygholami} :
\begin{align}
	h^a_{\;\;\mu} = Diag\left[1, a(t)\,\left(1-k\,r^2\right)^{-1/2},\,a(t)\,r,\, a(t)\,r\,\sin\theta\right].
\end{align}
{The} solution for spin-connection components will also be as follows \cite{preprint,coleylandrygholami} :
\begin{align}
	\omega_{122} =& \omega_{133} = \omega_{144} =  W_1(t),   &\omega_{234} =& -\omega_{243} = \omega_{342} = W_2(t), 
	\nonumber \\
	\omega_{233} =& \omega_{244} = - \frac{\sqrt{1-kr^2}}{a(t)r},  &\omega_{344} =&  \frac{\cos(\theta)}{a(t) r \sin(\theta)}, \label{Con:FLRW} 
\end{align}
where $W_1$ and $W_2$ depend on $k$-parameter and are defined as follows:
\begin{enumerate}
	\item $k=0$: $W_1=W_2=0$,
	\item $k=+1$: $W_1=0$ and $W_2(t)=\pm\,\frac{\sqrt{k}}{a(t)}$,	
	\item $k=-1$: $W_1(t)=\pm\,\frac{\sqrt{-k}}{a(t)}$ and $W_2=0$.		
\end{enumerate}

For any $W_1$ and $W_2$, we will obtain the same symmetric FEs set to solve for each subcase depending on the $k$-parameter. The previous coframe and spin-connection expressions were found by solving Equation \eqref{Intro:FS2} and imposing the null Riemann curvature condition (i.e., $R^a_{~b\mu\nu}=0$, as stated in ref. \cite{Coley:2019zld}). These solutions were also used in recent works on $F(T)$ TRW spacetime \cite{preprint,coleylandrygholami}. The FEs to be solved in the current paper are defined for each $k$-parameter case and will lead to different new teleparallel $F(T)$ solution classes. The FEs {\color{black} defined by Equations \eqref{1001a}--\eqref{1001b}} are also purely symmetric and valid on proper frames, as clearly showed in refs. \cite{preprint,coleylandrygholami}. {\color{black} More technically, Equation \eqref{1001b} is trivially satisfied and we will solve Equation \eqref{1001a} for each $k$-parameter.}

\subsection{Scalar Field Source Conservation Law Solutions}

{The scalar field source Lagrangian density $\mathcal{L}_{Source}$ term is defined as follows} \cite{Bahamonde:2021gfp,coleylandrygholami,scalar2,FTBcosmogholamilandry,scalarfieldKS}:
\begin{align}\label{300aa}
	\mathcal{L}_{Source} = \frac{h}{2}{\overset{\ \circ}{\nabla}}\,_{\nu}\phi\,\overset{\ \circ}{\nabla}\,^{\nu}\phi -h\,V\left(\phi\right).
\end{align}
{Equation} \eqref{300aa}, for a cosmological-like spacetime and a time-dependent $\phi=\phi(t)$ scalar field source, is as follows \cite{FTBcosmogholamilandry,scalarfieldKS}:
\begin{align}\label{300a}
	\mathcal{L}_{Source} = h\frac{\dot{\phi}^2}{2} -h\,V\left(\phi\right) ,
\end{align}
where $\dot{\phi}=\partial_t\,\phi$. By applying the least-action principle to Equation \eqref{300a}, the energy momentum tensor $T_{ab}$ is as follows \cite{hawkingellis1,coley03,nonvacSSpaper,nonvacKSpaper,roberthudsonSSpaper,cosmofluidsbohmer}: 
\begin{align}\label{1001d}
	T_{ab}= \left(P_{\phi}+\rho_{\phi}\right)\,u_a\,u_b+g_{ab}\,P_{\phi},
\end{align}
where $u_a=(-1,\,0,\,0,\,0)$ and the pressure $P_{\phi}$ and density $\rho_{\phi}$ are \cite{coleylandrygholami,scalarfieldKS}
\begin{align}\label{300b}
	P_{\phi}= \frac{\dot{\phi}^2}{2}-V\left(\phi\right) \quad \text{and}\quad   \rho_{\phi}=\frac{\dot{\phi}^2}{2}+V\left(\phi\right), 
\end{align}
with $\dot{\phi}=\phi_t$ and $V=V\left(\phi\right)$. The TRW spacetime scalar field conservation law is \cite{coleylandrygholami,FTBcosmogholamilandry}
\begin{align}
	3H\,\dot{\phi}+\ddot{\phi}+\frac{dV}{d\phi} =0, \label{302d}
\end{align}
{\color{black} where $H=\frac{\dot{a}}{a}$ is the Hubble parameter.} Equation \eqref{302d} is valid for any $\phi(t)$ scalar field expression {\color{black} and yields to a scalar potential $V\left(\phi(t)\right)$}. For the coming steps, we will solve the FEs by solving and satisfying Equation \eqref{302d}.

To complete the discussion on physical implications, from Equation \eqref{300b} and making the parallel between scalar field $\phi$ and the DE linear perfect fluid EoS equivalent {$P_{\phi}=\alpha_Q\rho_{\phi}$,} we need to define the quintessence coefficient index (or DE index) $\alpha_Q$ {as follows \cite{steinhardt1,steinhardt2,steinhardt3,steinhardt2024,scalarfieldKS}:}
\begin{align}\label{Quintessenceindex}
	\alpha_Q =& \frac{P_{\phi}}{\rho_{\phi}} = \frac{\dot{\phi}^2-2V\left(\phi\right)}{\dot{\phi}^2+2V\left(\phi\right)},
\end{align}
where the usual {{types} of DE} are defined as follows:
\begin{enumerate}
	\item \textbf{Quintessence} ${\bf -1<}\alpha_Q {\bf <-\frac{1}{3}}$: This describes a controlled accelerating universe expansion where energy conditions are always satisfied, i.e., $P_{\phi}+\rho_{\phi} > 0$ \cite{steinhardt1,steinhardt2,steinhardt3,steinhardt2024}. 
	
	\item \textbf{Phantom energy} $\alpha_Q {\bf <-1}$: This usually describes an uncontrolled universe expansion accelerating toward a Big Rip event \cite{farnes,baumframpton,caldwell1}. The energy condition is violated, i.e., $P_{\phi}+\rho_{\phi} \ngeq 0$.
	
	\item \textbf{Cosmological constant} $\alpha_Q{\bf =-1}$: This is an intermediate limit between the two previous and main types of DE, where $P_{\phi}+\rho_{\phi} = 0$. A constant scalar field source $\phi=\phi_0$ will directly lead to this case, according to Equation \eqref{Quintessenceindex}.

	\item \textbf{Quintom models:} This is a mixture of previous DE types, usually described by some double scalar field models \cite{quintom1,quintom2,quintom3,quintom4,quintomcoleytot}.
\end{enumerate}

By using Equation \eqref{Quintessenceindex}, we can find the perfect fluid equivalent for any new teleparallel $F(T)$ solution, potential $V\left(\phi\right)$ or ansatz. Equation \eqref{Quintessenceindex} is useful for making new teleparallel $F(T)$ solution classifications in terms of DE quintessence, phantom and quintom processes in cosmological model studies within the teleparallel framework.

%\newpage

\section{\boldmath$k=0$ Cosmological Solutions}\label{sect3}

The FEs and torsion scalar expressions for $\rho=\rho_{\phi}$ and $P=P_{\phi}$ defined by {Equation \eqref{300b}} are \cite{coleylandrygholami}:
%\begin{subequations}
\begin{align} 
	\kappa\left(\frac{\dot{\phi}^2}{2}+V\left(\phi\right)\right)=&-\frac{F(T)}{2}+6H^2F_T, \label{400a}
	\\
	\kappa(\dot{\phi}^2-V\left(\phi\right))=&\, \frac{F(T)}{2}-3\left(\dot{H}+2H^2 \right)F_T-3H\dot{F}_T , \label{400b}
	\\
	T=&\,6H^2.\label{400c}
\end{align}
%\end{subequations}
{By} 
merging Equations \eqref{400a} and \eqref{400b}, and then using Equation \eqref{400c}, we find a unified FE:
\begin{align} 
	-\frac{\sqrt{6}\,\kappa\dot{\phi}^2}{2}=& \,\partial_t\,\left({\sqrt{T}\,F_T}\right) . \label{401}
\end{align}
{For}  
any $\phi=\phi(t(T))$ scalar field definition and any potential $V(\phi)$ expression, we will use the $a(t)=a_0\,t^n$ and Equation \eqref{400c} to find that $t(T)=\frac{\sqrt{6}\,n}{\sqrt{T}}$. For any scalar field $\phi=\phi(t(T))$ and potential $V(\phi)$ expressions, the \textbf{general ${F(T)}$ solution formula} will be as follows:
\begin{align}\label{402}
	F(T) = -\Lambda_0+B\,\sqrt{T}-\frac{\sqrt{6}\,\kappa}{2}\,\left[\int\,dT'\,T'^{-1/2}\,\left[\int_{t(T')}\,dt'\,\dot{\phi}^2(t') \right]\right].
\end{align}
{The}  
first two terms of Equation \eqref{402} are identical to those from the linear perfect fluid $k=0$ TRW $F(T)$ solution found in ref. \cite{coleylandrygholami}. The last term of Equation \eqref{402} is clearly the scalar field source contribution.

We can compute the $F(T)$ solutions by using Equation \eqref{402} and the $a(t)=a_0\,t^n$ ansatz, yielding $t(T)=\frac{\sqrt{6}\,n}{\sqrt{T}}$ for the scalar field cases:
\begin{enumerate}
	\item \textbf{Power law general:} For $\phi(t){\bf =p_0\,t^p}$ and $p\neq \frac{1}{2}$, we find
	\begin{align}\label{403}
		F(T) = -\Lambda_0+B\,\sqrt{T}+\frac{\sqrt{6}\,\kappa\,p_0^2\,p^2(6n^2)^{p-\frac{1}{2}}}{2(2p-1)(p-1)}\,T^{1-p}.
	\end{align}
	
	\item \textbf{Power law special:} For $\phi(t){\bf =p_0\,t^{\frac{1}{2}}}$, we find
	\begin{align}\label{404}
		F(T) = -\Lambda_0+B\,\sqrt{T}+\frac{\sqrt{6}\,\kappa\,p_0^2}{8}\,\sqrt{T}\,\ln\,(T).
	\end{align}
	
	\item \textbf{Logarithmic:} For a field defined as $\phi(t){\bf =p_0\,\ln\left(t\right)}$, we find
	\begin{align}\label{405}
		F(T) = -\Lambda_0+B\,\sqrt{T}+\frac{\kappa\,p_0^2}{2n}\,T.
	\end{align}

	\item \textbf{Exponential:} For $\phi(t){\bf =p_0\,\exp\left(p\,t\right)}$, we find
	\vspace{-12pt}
	
%	\begin{adjustwidth}{-\extralength}{0cm}
		%\centering %% If there is a figure in wide page, please release command \centering
		\begin{align}\label{406}
			F(T) = -\Lambda_0+B\,\sqrt{T}-\,\frac{\sqrt{6}\kappa}{2}\,p_0^2\,p \left[2\sqrt{6}\, n p \,{Ei}_{1}\! \left(-\frac{2 p \sqrt{6}\, n}{\sqrt{T}}\right)+{\sqrt{T}}\, \exp\left(\frac{2 p \sqrt{6}\, n}{\sqrt{T}}\right)\right],
		\end{align}
%	\end{adjustwidth}
	where ${Ei}_{1}\!(x)$ is an exponential integral function.

\end{enumerate}

There are several other possible $\phi(t)$ source terms which may lead to {\color{black} additional new teleparallel $F(T)$ solutions by using} Equation \eqref{402}'s general formula.

{\color{black}
	Therefore, there are some cosmological significations concerning the $n$-parameter in the $a(t)=a_0\,t^n$ ansatz expression. This last parameter in the scale factor $a(t)$ determines the type of universe evolution as a physical system \cite{coleylandrygholami,FTBcosmogholamilandry}:
	\begin{enumerate}
		\item ${\bf n<0}$: Contracting universe. This is a "Big Crunch"-type scenario for a large $|n|$-parameter scenario.
		
		\item ${\bf n=0}$: Static universe. This is the limit between expanding and contracting universe scenarios.
		
		\item ${\bf 0<n<1}$: This is slow and controlled universe expansion, but a non-inflationary scenario. 
		
		\item ${\bf n=1}$: This is linear universe expansion and the limit between slow and fast universe expansion scenarios.
		
		\item ${\bf 1<n<\infty}$: The is fast, inflationary and controlled universe expansion. This is a plausible dark energy quintessence case because it is an inflationary scenario.
		
		\item ${\bf n\gg 1}$ or ${\bf n\,\rightarrow\,\infty}$: This is very fast and uncontrolled universe expansion. This strong inflationary case is so far the best phantom dark energy scenario description and leads to "Big Rip" singularity after a determined time delay.
	\end{enumerate}
	
	For $F(T)$ FEs solutions for Equations \eqref{403}--\eqref{405}, we see that Equation \eqref{404} is independent of the $n$-parameter and can be considered a reference solution. Figure \ref{figure1} compares several $n>0$ expanding universe scenarios for different types of scalar field source. Each subfigure $F(T)$ plot has a similar form for any values of $n$ and a continuum arises for the $k=0$ case. Equation \eqref{406} $F(T)$ solutions for exponential sources are very different for each value of $n$, compared to the other sources, and this is the main reason for non-plotting. This scalar field source constitutes a very special case and some precautions should be taken, as this case is possible in future studies. All the coming non-zero $k$-parameter $F(T)$ solutions will be computed by thinking in terms of slow ($n=\frac{1}{2}$), linear ($n=1$), fast ($n=2$) and very fast ($n\gg 1$ or $n=20$ in practical) universe expansion case~scenarios. 
	
	\vspace{-15pt}
	
	\begin{figure} %[H]
		\hspace{-17pt}\includegraphics[scale=0.79]{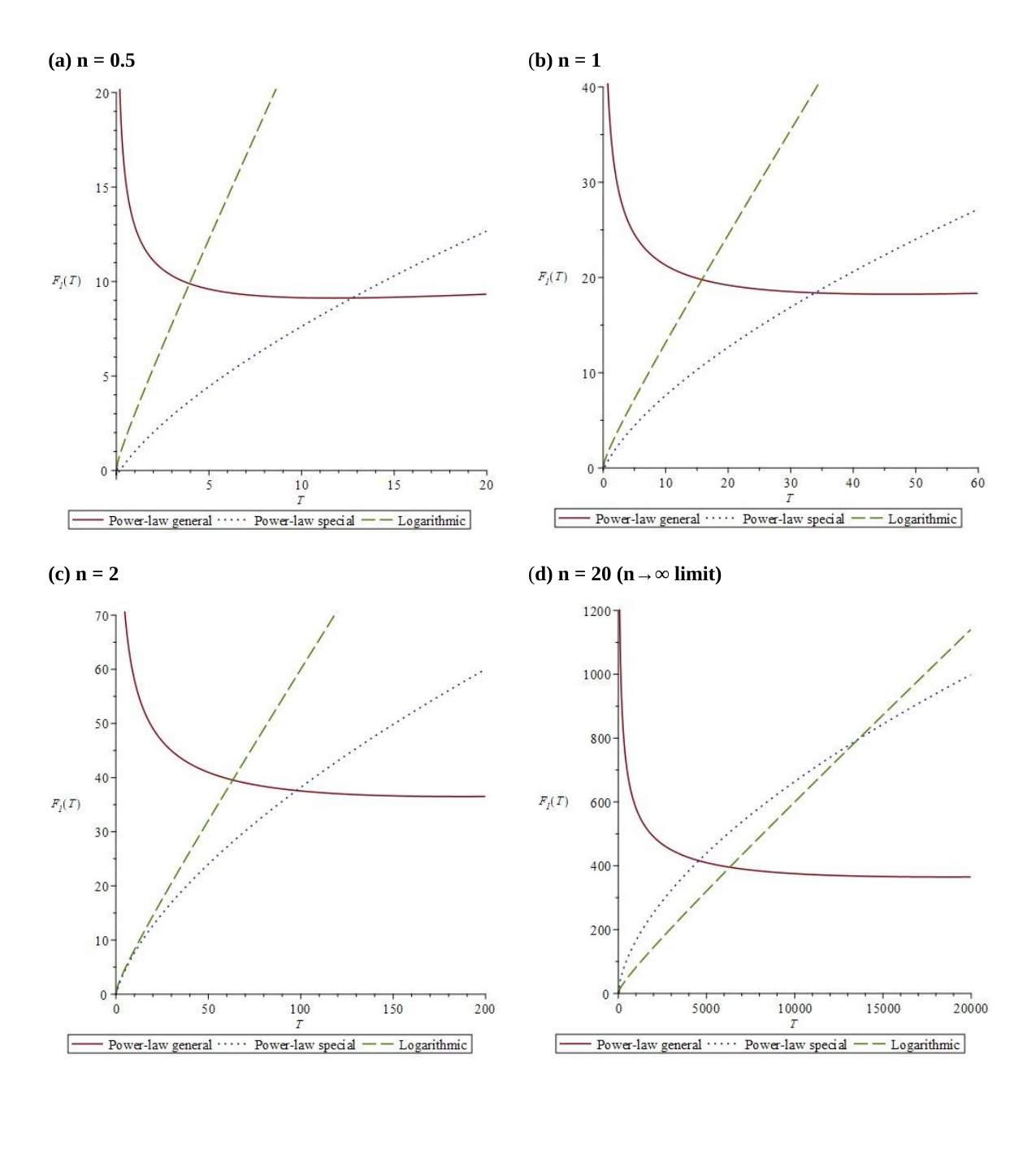}
		\vspace{-50pt}
		
		\caption{{\color{black}{Plots} %MDPI: 1. Please use periods as decimal signs instead of commas, e.g., "0,1" should be "0.1". 2. Please use commas to separate thousands for numbers with five or more digits (not four digits) in the picture, e.g., "10000" should be "10,000". AL: OK on all figures.
				of $F_1(T)=\frac{F(T)+\Lambda_0}{B}$ versus torsion scalar $T$, described by Equations \eqref{403}--\eqref{405}, for different types of scalar field sources and $n$-parameter with $\frac{\kappa\,p_0^2}{2B}=1$ and $p=1.3$. Note that the $n=20$ case in subfigure (\textbf{d}) represents Equations \eqref{403}--\eqref{405} for the $n\,\rightarrow\,\infty$ limit.}}
		\label{figure1}
	\end{figure}
	
	\vspace*{0.4cm}
	
	However, the $F(T)$ solution can be compared with some flat cosmological observation data by fitting those over the Figure \ref{figure1} curves to determine which are the most optimal parameters and scalar field sources, as was carried out in refs. \cite{dixit,ftbcosmo3}. We should note that data fitting was already suggested in ref. \cite{FTBcosmogholamilandry} as a possible future work. We can also extend this type of work to a more thermodynamic-based study on cosmological quantities, as mentioned in ref. \cite{FTBcosmogholamilandry}.

	For the other possible $a(t)$ ansatz, there are some limited choices. For example, the pure exponential $a(t)=a_0\,\exp\left(n\,t\right)$ will lead to a constant Hubble parameter $H=H_0=n$ and a constant torsion scalar $T=6\,H_0^2=6\,n^2$, a teleparallel de Sitter (TdS) spacetime and also a general relativity (GR) solution \cite{TdSpaper,coleylandrygholami}.  
	
	%\newpage

}

%\newpage

\section{\boldmath$k=-1$ Cosmological Solutions}\label{sect4}

The FEs and torsion scalar for $\rho=\rho_{\phi}$ and $P=P_{\phi}$, defined by Equation \eqref{300b}, are as follows \cite{coleylandrygholami}:
%\begin{subequations}
\begin{align}%\fl 
	\kappa\left(\frac{\dot{\phi}^2}{2}+V\left(\phi\right)\right)=&-\frac{F(T)}{2}+6H\left(H+\frac{\delta\sqrt{-k}}{a}\right)\,F_T, \label{500a}
	\\
	%\fl
	\kappa(\dot{\phi}^2-V\left(\phi\right))=&\frac{F(T)}{2}-3\left(\dot{H}+2H^2+2H\frac{\delta\sqrt{-k}}{a}-\frac{k}{a^2} \right)\,F_T-3\left(H+\frac{\delta\sqrt{-k}}{a}\right)\,\dot{F}_T, \label{500b}
	\\
	T=&6\left( H+ \frac{\delta\sqrt{-k}}{a}\right)^2. \label{500c}
\end{align}
%\end{subequations}
{By}  
merging Equations \eqref{500a} and \eqref{500b}, the unified FE will be expressed as follows:
\begin{align}%\fl 
	-\frac{\kappa\dot{\phi}^2}{2}=& \left(\dot{H}-\frac{k}{a^2} \right)\,F_T+\left(H+\frac{\delta\sqrt{-k}}{a}\right)\,\dot{F}_T. \label{501}
\end{align}
{The}  
general $F(T)$ solution to be computed, from Equation \eqref{501}, will be expressed as follows:
\begin{align}\label{502}
	F(T)=&\,-\Lambda_0+\int\,dT\Bigg[\Bigg(C_1-\frac{\kappa}{2}\int_{t(T)}\,dt'\,\exp\left(\int_{t'}\,dt''\,\left(\frac{a^2\,\dot{H}-k}{a\,\left(a\,H+\delta\sqrt{-k}\right)} \right)\right)
	\nonumber\\
	&\,\times\,\frac{a\,\dot{\phi}^2(t')}{\left(a\,H+\delta\sqrt{-k}\right)}\Bigg)\exp\left(-\int_{t(T)}\,dt'\,\left(\frac{a^2\,\dot{H}-k}{a\,\left(a\,H+\delta\sqrt{-k}\right)} \right)\right)\Bigg] .
\end{align}
{By}  
applying a power law ansatz $a(t)=a_0\,t^n$ to Equation \eqref{502}, the $F(T)$ solution will simplify as follows:
\begin{align}\label{503}
	F(T)=&\,-\Lambda_0+\int\,dT\Bigg[\Bigg(C_1-\,\kappa\int_{t(T)}\,dt'\,\dot{\phi}^2(t')\,\exp\left(-\frac{\delta\,\sqrt{-k}}{a_0\,(n-1)}\,t'^{1-n}\right)\Bigg)
	\nonumber\\
	&\,\times\,\frac{\exp\left(\frac{\delta\,\sqrt{-k}}{a_0\,(n-1)}\,t^{1-n}(T)\right)}{2nt^{-1}(T)+2\frac{\delta\,\sqrt{-k}}{a_0}\,t^{-n}(T)} \Bigg] .
\end{align}
{From}  
Equation \eqref{500c}, the characteristic equation leading to $t(T)$ expressions is
\begin{align}\label{504}
	0=\delta_1\sqrt{\frac{T}{6}}- n\,t^{-1}- \frac{\delta\sqrt{-k}}{a_0}\,t^{-n}.
\end{align}
{There}  
are specific values of $n$ leading to Equation \eqref{504}-based analytical teleparallel $F(T)$ solutions:
\begin{enumerate}
	\item ${\bf n=\frac{1}{2}}$: Equation \eqref{504} becomes
	\begin{align}\label{505}
		0=& t^{-1}+ \frac{2\delta\sqrt{-k}}{a_0}\,t^{-\frac{1}{2}}-\delta_1\sqrt{\frac{2T}{3}},
		\nonumber\\
		\Rightarrow &\quad t^{-\frac{1}{2}}(T)= -\frac{\delta\sqrt{-k}}{a_0}+ \delta_2\sqrt{-\frac{k}{a_0^2}+\delta_1\sqrt{\frac{2T}{3}}} ,
	\end{align}	
	where $\delta_2=\pm 1$. Then, Equation \eqref{503} for the $F(T)$ solution is
	\begin{align}\label{506}
		F(T)=&\,-\Lambda_0+\int\,dT\Bigg[\Bigg(C_1-{\kappa\,\delta_1\,\sqrt{\frac{3}{2}}}\int_{t(T)}\,dt'\,\dot{\phi}^2(t')\,\exp\left(\frac{2\delta\,\sqrt{-k}}{a_0}\,t'^{\frac{1}{2}}\right)\Bigg)
		\nonumber\\
		&\,\times\,T^{-1/2}\,\exp\left(2\,\left[1- \delta_2\sqrt{1-\frac{a_0^2\,\delta_1}{k}\sqrt{\frac{2T}{3}}} \right]^{-1}\right)\Bigg] .
	\end{align}
	{Equation}  
	\eqref{506}, by setting $\frac{\delta  \sqrt{-k}}{a_0}=1$ and a power-law scalar field $\phi(t)=p_0\,t^p$, yields new analytical $F(T)$ solutions for the following subcases:
	\begin{enumerate}
		\item ${\bf p= \frac{3}{4}}$:
		\small
		\begin{align}\label{506a}
			F(T)=&\,-\Lambda_0+C_1\, \exp\left(2\,\left[1- \delta_2\sqrt{1+\delta_1\sqrt{\frac{2T}{3}}} \right]^{-1}\right)\left[1- \delta_2\sqrt{1+\delta_1\sqrt{\frac{2T}{3}}} \right]^{2}
			\nonumber\\
			&\,\quad-\kappa\,\delta_1 p_0^2\,\frac{9}{8}\,\sqrt{\frac{3}{2}}\,T^{1/2} .
		\end{align}	
		\normalsize	
		
		\item${\bf p=1}$:
		\small
		\begin{align}\label{506b}
			F(T)=&\,-\Lambda_0+C_1\exp\left(2\,\left[1- \delta_2\sqrt{1+\delta_1\sqrt{\frac{2T}{3}}} \right]^{-1}\right)\left[1- \delta_2\sqrt{1+\delta_1\sqrt{\frac{2T}{3}}} \right]^{2}+\kappa\, p_0^2\,
			\nonumber\\
			&\,\times\, \Bigg(\delta_1\,\sqrt{\frac{3}{2}}\,T^{1/2}-2 \delta_{2} \sqrt{9+3 \delta_{1} \sqrt{6}\,{\sqrt{T}}}+3 \ln\! \left(\frac{1+\delta_{2} \sqrt{1+\delta_1\sqrt{\frac{2T}{3}}}}{1-\delta_{2} \sqrt{1+\delta_1\sqrt{\frac{2T}{3}}}}\right)
			\nonumber\\
			&\,-\frac{3}{2} \ln\! \left(-2 T\right)\Bigg).
		\end{align}
		\normalsize
		
	\end{enumerate}
	There are several other possible $F(T)$ solutions; these can be obtained by setting other values of $\frac{\delta  \sqrt{-k}}{a_0}$, $p$ and/or other scalar field $\phi(t)$ expressions inside the general expression of Equation \eqref{506}. However, we can expect that such cases will yield more significant expressions of $F(T)$.

	\item ${\bf n=1}$: Equation \eqref{504} becomes
	\begin{align}\label{507}
		0=& \delta_1\sqrt{\frac{T}{6}}- \left(1+ \frac{\delta\sqrt{-k}}{a_0}\right)\,t^{-1},
		\nonumber\\
		\Rightarrow &\quad t(T)=\delta_1\left(1+ \frac{\delta\sqrt{-k}}{a_0}\right)\frac{\sqrt{6}}{\sqrt{T}} ,
	\end{align}
	Equation \eqref{503} for the $F(T)$ solution becomes
	\begin{align}\label{508}
		F(T)=&\,-\Lambda_0+B\,T^{\frac{\delta\,\sqrt{-k}}{2a_0}+\frac{1}{2}}
		\nonumber\\
		&\quad-\frac{\delta_1\sqrt{6}\kappa}{2\left(\delta_1\sqrt{6}\left(1+ \frac{\delta\sqrt{-k}}{a_0}\right)\right)^{\frac{\delta\,\sqrt{-k}}{a_0}}}\int\,dT\,T^{\frac{\delta\,\sqrt{-k}}{2a_0}-\frac{1}{2}}\left[\int_{t(T)}\,dt'\,\dot{\phi}^2(t')\,t'^{\frac{\delta\,\sqrt{-k}}{a_0}}\right].
	\end{align}
	Equation \eqref{508} yields new analytical $F(T)$ solutions for the following cases:
	\begin{enumerate}
		\item \textbf{General Power law} $\phi(t){\bf = p_0\,t^p}$:
		\begin{align}\label{508a}
			F(T)=&\,-\Lambda_0+B\,T^{\frac{\delta\,\sqrt{-k}}{2a_0}+\frac{1}{2}}+\frac{\delta_1\sqrt{6}\kappa p_0^2\,p^2\left[\delta_1\left(1+ \frac{\delta\sqrt{-k}}{a_0}\right)\sqrt{6}\right]^{2p-1}}{2\,\left(2p-1+\frac{\delta\,\sqrt{-k}}{a_0}\right)(p-1)}\,T^{1-p}.
		\end{align}
		\item \textbf{Special Power law} $\phi(t){\bf = p_0}\,t^{{\bf \frac{1}{2}}-\frac{\delta\,{\bf \sqrt{-k}}}{{\bf 2a_0}}}$:
		\begin{align}\label{508b}
			F(T)=&\,-\Lambda_0+B\,T^{\frac{\delta\,\sqrt{-k}}{2a_0}+\frac{1}{2}}+\frac{\kappa p_0^2 \left(1-\frac{\delta\sqrt{-k}}{a_0}\right)^2\,\, T^{\frac{\delta\,\sqrt{-k}}{2a_0}+\frac{1}{2}}  \ln\! \left(T\right)}{8\left(\delta_1\sqrt{6}\right)^{\frac{\delta\,\sqrt{-k}}{a_0}-1}\left(1+ \frac{\delta\sqrt{-k}}{a_0}\right)^{\frac{\delta\,\sqrt{-k}}{a_0}+1}}.
		\end{align}
		
		\item \textbf{Logarithmic} $\phi(t){\bf = p_0\ln\left(pt\right)}$:
		\begin{align}\label{508c}
			F(T)=&\,-\Lambda_0+B\,T^{\frac{\delta\,\sqrt{-k}}{2a_0}+\frac{1}{2}}+\frac{\kappa\,p_0^2}{2\left(1+ \frac{k}{a_0^2}\right)}\,T.
		\end{align}
		
		\item \textbf{Exponential} $\phi(t){\bf = p_0\exp\left(pt\right)}$:
		\begin{align}\label{508d}
			F(T)=&\,-\Lambda_0+B\,T^{\frac{\delta\,\sqrt{-k}}{2a_0}+\frac{1}{2}}-\frac{\delta_1\sqrt{6}\kappa\,p_0^2\,p^2}{2\left(\delta_1\sqrt{6}\left(1+ \frac{\delta\sqrt{-k}}{a_0}\right)\right)^{\frac{\delta\,\sqrt{-k}}{a_0}}}
			\nonumber\\
			&\quad \times \,\int\,dT\,T^{\frac{\delta\,\sqrt{-k}}{2a_0}-\frac{1}{2}}\left[\int_{t(T)}\,dt'\,\exp \left(2p\,t'\right) \,t'^{\frac{\delta\,\sqrt{-k}}{a_0}}\right].
		\end{align}
		There is no general solution for Equation \eqref{508d}. However, there are specific solutions:
		\begin{itemize}
			\item $\frac{\delta\,\sqrt{-k}}{a_0}{\bf =1}$:
			\begin{align}\label{508da}
				F(T)=&\,-\Lambda_0+B\,T
				\nonumber\\
				&\quad -\frac{\kappa\,p_0^2}{16}\Bigg[96 \,{Ei}_{1}\! \left(-\frac{4 p \delta_1  \sqrt{6}}{\sqrt{T}}\right) p^{2} -{e}^{\frac{4 p \delta_1  \sqrt{6}}{\sqrt{T}}} \left(T-4 {\sqrt{T}}\, p \delta_1  \sqrt{6}\right)\Bigg].
			\end{align}
			\item $\frac{\delta\,\sqrt{-k}}{a_0}{\bf =2}$:
			\begin{align}\label{508db}
				F(T)=&\,-\Lambda_0+B\,T^{\frac{3}{2}}-\frac{\delta_1 \kappa\,p_0^2}{18p} \,\Bigg[108 \delta_1 p^{3} {Ei}_{1}\! \left(-\frac{6 p \delta_1  \sqrt{6}}{\sqrt{T}}\right)
				\nonumber\\
				&\,+{e}^{\frac{6 p \delta_1  \sqrt{6}}{\sqrt{T}}} \left(\frac{T^{\frac{3}{2}} \sqrt{6}}{36}+3 \sqrt{6}\,p^2 {\sqrt{T}}- \delta_1\,p\,T \right) \Bigg].
			\end{align}
			
		\end{itemize} 
		There are several possible new $F(T)$ solutions for other values of $\frac{\delta\,\sqrt{-k}}{a_0}$ and using Equation \eqref{508d}. 
	\end{enumerate}
	There are additional possible new $F(T)$ solutions from the Equation \eqref{508} integral with other types of scalar field sources.
	
	\item ${\bf n=2}$: Equation \eqref{504} becomes
	\begin{align}\label{509}
		0=& t^{-2}+ \frac{2\delta a_0}{\sqrt{-k}}\,t^{-1}-\delta_1 \delta a_0\sqrt{-\frac{T}{6k}},
		\nonumber\\
		\Rightarrow &\quad t^{-1}(T)= -\frac{\delta a_0}{\sqrt{-k}}+\delta_2\sqrt{\delta_1 \delta a_0\sqrt{-\frac{T}{6k}}-\frac{a_0^2}{k}},
	\end{align}
	where $\delta_2=\pm 1$. Equation \eqref{503} becomes
	\begin{align}\label{510}
		F(T)=&\,-\Lambda_0+\int\,dT\Bigg[\Bigg(C_1-\frac{\delta_1\,\kappa\,\sqrt{6}}{2}\int_{t(T)}\,dt'\,\dot{\phi}^2(t')\,\exp\left(-\frac{\delta\,\sqrt{-k}}{a_0}\,t'^{-1}\right)\Bigg)
		\nonumber\\
		&\,\times\,\frac{\exp\left[-1+\delta_2\sqrt{1+\frac{\delta_1\,\delta}{a_0}\sqrt{\frac{-k}{6}}\sqrt{T}}\right]}{\sqrt{T}} \Bigg] .
	\end{align}
	By using a power-law scalar field $\phi(t)=p_0\,t^p$ and setting $\frac{\delta\,\sqrt{-k}}{a_0}=1$, Equation \eqref{510} yields new analytical $F(T)$ solutions for some $p<0$ subcases:
	\begin{enumerate}
		\item  ${\bf p=-\frac{1}{2}}$:
		\begin{align}\label{510a}
			F(T)=&\,-\Lambda_0+C_1 \,\left[-1+\delta_2\sqrt{1+\delta_1\,\sqrt{\frac{T}{6}}}\right] \exp\left[-1+\delta_2\sqrt{1+\delta_1\,\sqrt{\frac{T}{6}}}\right]
			\nonumber\\
			&\quad\,-\delta_2\,\kappa\,p_0^2\,\left(1+\delta_1\,\sqrt{\frac{T}{6}}\right)^{3/2}  .
		\end{align}

		\item ${\bf p=-1}$:
		\begin{align}\label{510b}
			F(T)=&\,-\Lambda_0+C_1 \,\left[-1+\delta_2\sqrt{1+\delta_1\,\sqrt{\frac{T}{6}}}\right] \exp\left[-1+\delta_2\sqrt{1+\delta_1\,\sqrt{\frac{T}{6}}}\right]
			\nonumber\\
			&\quad\, -\kappa\,p_0^2\,\left(2\sqrt{6}\,\delta_1\,T^{1/2}+\frac{T}{2}\right) .
		\end{align}

	\end{enumerate}
	There are several other possible $F(T)$ solutions by setting other values of $\frac{\delta  \sqrt{-k}}{a_0}$, $p$ and/or other scalar field $\phi(t)$ expressions. However, we can expect that such cases will yield a more significant expression of $F(T)$.

	\item ${\bf n\gg 1}$: Equation \eqref{504} {\color{black} leads to $t(T)= n\,\sqrt{\frac{6}{T}}$ with $\delta_1=1$. From this point, we obtain Equation \eqref{402}'s exact formula and then Equations \eqref{403}--\eqref{406} as $F(T)$ solutions under the large $n$ limit. The $n\gg 1$ solutions are the same as for the $k=0$ flat cosmological case and the graphs are shown in Figure \ref{figure1}d.}

\end{enumerate}

%\newpage
{\color{black}
	We can also compare the $n=\frac{1}{2}$, $1$, $2$ and large $n$ expanding universe scenarios by plotting the $F(T)$ solutions for these values of $n$, as shown in Section \ref{sect3}. Figure \ref{figure2} shows various types of scalar field source. Figure \ref{figure2}a,c use the power law ansatz for specific values of $p$ and essentially show a comparison between the different values of $p$ for power law scalar field sources. As for the Section \ref{sect3} flat cosmological case, we can also suggest comparing the curves in Figure \ref{figure2} with observational data, as carried out in refs. \cite{dixit,ftbcosmo3} by using some techniques such as the tensor-to-scalar ratio, to name only one method. We can again suggest, for this type of future work, a thermodynamic-based parameter evolution study, as suggested in ref. \cite{FTBcosmogholamilandry}.
	
	\vspace{-15pt}
	\begin{figure} %[H]
		\hspace{-13pt}\includegraphics[scale=0.79]{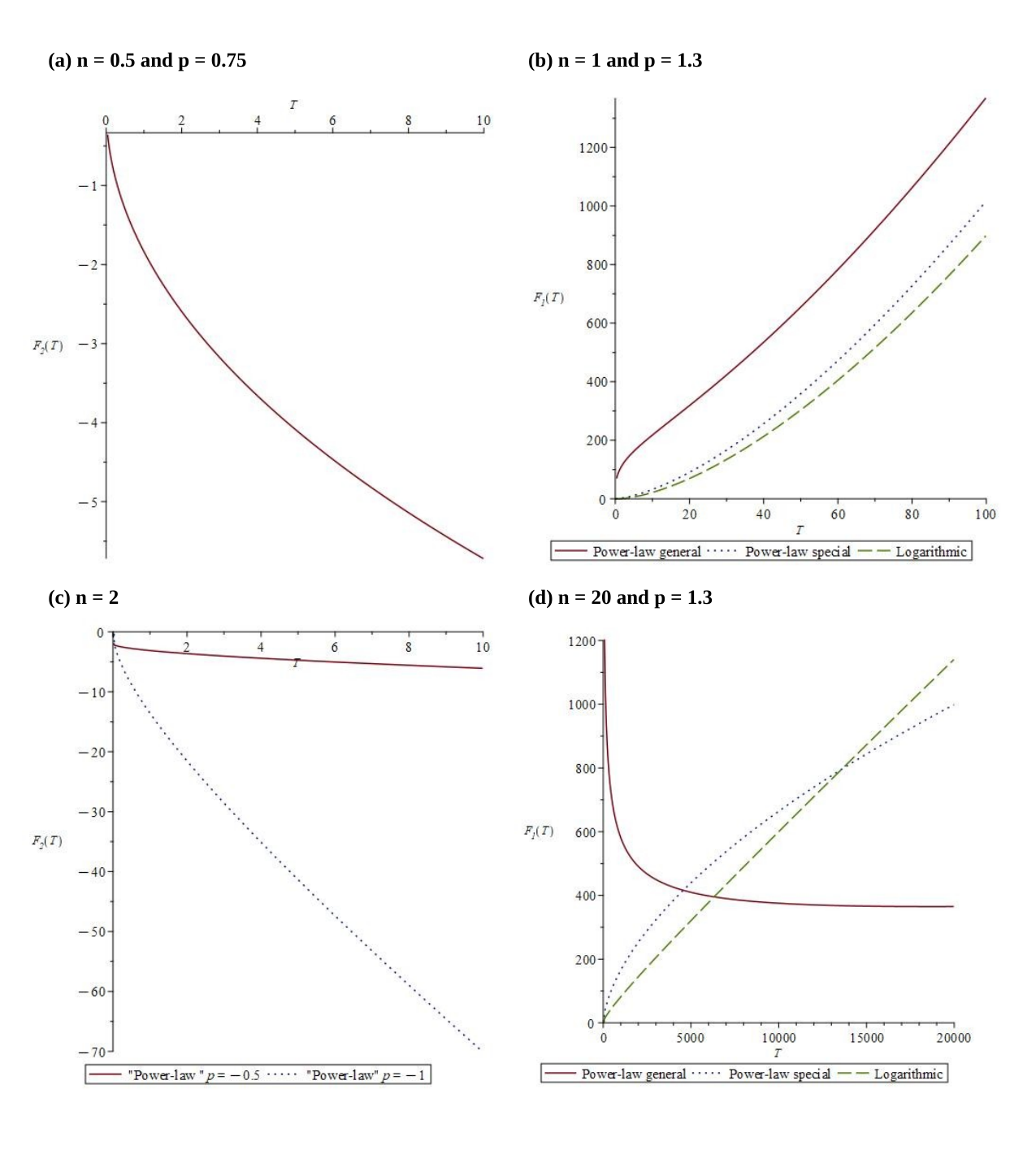}
		\vspace{-36pt}
		\caption{{\color{black}{Plots} %MDPI: 1. Please use periods as decimal signs instead of commas, e.g., "0,1" should be "0.1". 2. Please use commas to separate thousands for numbers with five or more digits (not four digits) in the picture, e.g., "10000" should be "10,000". AL: OK it is fixed on all figures.
				of $F_2(T)=F(T)+\Lambda_0$ versus torsion scalar $T$ described by the $F(T)$ from Equations \eqref{506a}--\eqref{510b} for different types of scalar field sources, $n$ and $p$ parameters with $\frac{\kappa\,p_0^2}{2}=1$ and $\frac{\delta\,\sqrt{-k}}{a_0}=2$. Note that the $n=20$ case in subfigure (\textbf{d}) represents Equations \eqref{403}--\eqref{405} for the $n\,\rightarrow\,\infty$ limit, where $B=1$ and $\delta_1=1$, which is exactly the same in Figure \ref{figure1}d when $B=1$ in $F_2(T)$.}}
		\label{figure2}
	\end{figure}
	
}

\newpage

\section{\boldmath$k=+1$ Cosmological Solutions}\label{sect42}

The FEs and torsion scalar for $\rho=\rho_{\phi}$ and $P=P_{\phi}$, defined by Equation \eqref{300b}, are as follows \cite{coleylandrygholami}:
%\begin{subequations}
\begin{align}
	\kappa\left(\frac{\dot{\phi}^2}{2}+V\left(\phi\right)\right)=&-\frac{F(T)}{2}+6H^2\,F_T, \label{550a}
	\\
	\kappa(\dot{\phi}^2-V\left(\phi\right))=& \frac{F(T)}{2}-3\left(\dot{H}+2H^2-\frac{k}{a^2} \right)\,F_T-3H\,\dot{F}_T, \label{550b}
	\\
	T=& 6\left[ H^2- \frac{k}{a^2}\right].  \label{550c}
\end{align}
%\end{subequations}
{By}  
merging Equations \eqref{550a} and \eqref{550b}, the unified FE will be expressed as follows:
\begin{align}
	-\frac{\kappa\dot{\phi}^2}{2}=&\left(\dot{H}-\frac{k}{a^2} \right)\,F_T+H\,\dot{F}_T, \label{551}
\end{align}
{The}  
general $F(T)$ solution to be computed from Equation \eqref{551} will be expressed as follows:
\vspace{-15pt}

%\begin{adjustwidth}{-\extralength}{0cm}
	%\centering %% If there is a figure in wide page, please release command \centering
	\begin{align}\label{552}
		F(T)=&\,-\Lambda_0+\int\,\frac{dT}{H(t(T))}\Bigg[C_1-\frac{\kappa}{2}\int_{t(T)}\,dt'\,\exp\left(-k\,\int_{t'}\,\frac{dt''}{H\,a^2}\right)\dot{\phi}^2(t')\Bigg]\exp\left(k\int_{t(T)}\,\frac{dt'}{H\,a^2}\right) .
	\end{align}
%\end{adjustwidth}
{Equation}  
\eqref{552} becomes, by applying the power-law ansatz $a(t)=a_0\,t^n$ to Equation \eqref{552},
\begin{align}\label{553}
	F(T)=&\,-\Lambda_0+\frac{1}{n}\int\,dT\,t(T)\,\Bigg[C_1-\frac{\kappa}{2}\int_{t(T)}\,dt'\,\dot{\phi}^2(t')\,\exp\left(-\frac{k\,t'^{2(1-n)}}{2n(1-n)\,a_0^2}\right)\Bigg]
	\nonumber\\
	&\quad \times\,\exp\left(\frac{k\,t^{2(1-n)}(T)}{2n(1-n)\,a_0^2}\right) .
\end{align}
{From}  
Equation \eqref{550c}, the characteristic equation leading to $t(T)$ expressions is as follows:
\begin{align}\label{554}
	0=& \frac{T}{6}-\frac{n^2}{t^2}+ \frac{k}{a_0^2}\,t^{-2n} .
\end{align}
{There}  
are specific values of $n$ leading to Equation \eqref{554}-based analytical teleparallel $F(T)$ solutions:
\begin{enumerate}
	\item ${\bf n=\frac{1}{2}}$: Equation \eqref{554} becomes
	\begin{align}\label{555}
		0=& t^{-2}- \frac{4k}{a_0^2}\,t^{-1}-\frac{2T}{3},
		\nonumber\\
		\Rightarrow &\quad t^{-1}(T)= \frac{2k}{a_0^2} +2\delta_2\sqrt{\frac{k^2}{a_0^4}+\frac{T}{6}} ,			 .
	\end{align}
	where $\delta_2=\pm 1$. Equation \eqref{553} becomes
	%\small
	\vspace{-15pt}
	
	%\begin{adjustwidth}{-\extralength}{0cm}
		%\centering %% If there is a figure in wide page, please release command \centering
		\small
		\begin{align}\label{556}
			F(T)=&\,-\Lambda_0+\int\,dT\,\Bigg[C_1-\frac{\kappa\,a_0^2}{2k}\int_{t(T)}\,dt'\,\dot{\phi}^2(t')\,\exp\left(-\frac{2k\,t'}{a_0^2}\right)\Bigg]\,\frac{\exp\left(4\left(1 +\delta_2\sqrt{1+\frac{a_0^4}{6k^2}\,T}\right)^{-1}\right)}{\left(1 +\delta_2\sqrt{1+\frac{a_0^4}{6k^2}\,T}\right)} .
		\end{align}
	%\end{adjustwidth}
		\normalsize
	Equation \eqref{556}, by setting $\frac{k}{a_0^2}=1$, yields new analytical $F(T)$ solutions for the scalar field:
	\begin{enumerate}
		\item \textbf{Linear Power law} $\phi(t){\bf =p_0\,t}$:
		\begin{align}\label{556a}
			F(T)=&\,-\Lambda_0+C_1\Bigg[\left(1+\delta_{2} \sqrt{1+\frac{T}{6}}\right)\, \exp\left(4\left(1 +\delta_2\sqrt{1+\frac{T}{6}}\right)^{-1}\right)
			\nonumber\\
			&\,+{Ei}_{1}\! \left(-4\left(1 +\delta_2\sqrt{1+\frac{T}{6}}\right)^{-1}\right)\Bigg] 
			\nonumber\\
			&+\frac{3\kappa\,p^2\,p_0^2}{2}\,\Bigg[2 \delta_{2} \sqrt{1+\frac{T}{6}}- \ln\! \left(T \right)+ \ln\! \left(\frac{\delta_{2} \sqrt{1+\frac{T}{6}}-1}{\delta_{2} \sqrt{1+\frac{T}{6}}+1}\right)\Bigg]  .
		\end{align}
		
		\item \textbf{Quadratic Power law} $\phi(t){\bf =p_0\,t^2}$:
		%\small
		\begin{align}\label{556c}
			F(T)=&\,-\Lambda_0+C_1\Bigg[\left(1+\delta_{2} \sqrt{1+\frac{T}{6}}\right)\, \exp\left(4\left(1 +\delta_2\sqrt{1+\frac{T}{6}}\right)^{-1}\right)
			\nonumber\\
			&\,+3{Ei}_{1}\! \left(-4\left(1 +\delta_2\sqrt{1+\frac{T}{6}}\right)^{-1}\right)\Bigg] 
			\nonumber\\
			&\,+\frac{9\kappa\,p^2\,p_0^2}{4\,T}\,\Bigg[ \frac{12}{T}-1+4\left(1-\frac{3}{T}\right)\delta_{2}\left(1+\frac{T}{6}\right)^{3/2}\Bigg]  .
		\end{align}		
		\normalsize
		
		\item \textbf{Exponential} $\phi(t){\bf =p_0\exp\left(pt\right)}$:
		\begin{align}\label{556b}
			F(T)=&\,-\Lambda_0+C_1\Bigg[\left(1+\delta_{2} \sqrt{1+\frac{T}{6}}\right)\, \exp\left(4\left(1 +\delta_2\sqrt{1+\frac{T}{6}}\right)^{-1}\right)
			\nonumber\\
			&\,+3{Ei}_{1}\! \left(-4\left(1 +\delta_2\sqrt{1+\frac{T}{6}}\right)^{-1}\right)\Bigg] 
			\nonumber\\
			&\,-\frac{3\kappa\,p^2\,p_0^2}{(p-1)}\,\Bigg[\left(1 +\delta_2\sqrt{1+\frac{T}{6}}\right)\, \exp\left(4p\left(1 +\delta_2\sqrt{1+\frac{T}{6}}\right)^{-1}\right) 
			\nonumber\\
			&\,+(4p-1) \,{Ei}_{1}\! \left(-4p\left(1 +\delta_2\sqrt{1+\frac{T}{6}}\right)^{-1}\right)\Bigg] .
		\end{align}
		
	\end{enumerate}
	There are several other possible $F(T)$ solutions that can be achieved by setting other values of $\frac{k}{a_0}$, $p$ and/or other scalar field $\phi(t)$ expressions. However, we can expect that such cases will yield a more significant expression of $F(T)$.

	\item ${\bf n=1}$: Equation \eqref{554} becomes
	\begin{align}\label{557}
		0=& T-6\left(1- \frac{k}{a_0^2}\right)\,t^{-2},
		\nonumber\\
		\Rightarrow &\quad t^2(T)=\frac{6\left(1- \frac{k}{a_0^2}\right)}{T} .
	\end{align}
	Equation \eqref{553} becomes
	\begin{align}\label{558}
		F(T)=&\,-\Lambda_0+B\,T^{\frac{1}{2}-\frac{k}{2a_0^2}}+\frac{\kappa}{2}\left(6\left(1- \frac{k}{a_0^2}\right)\right)^{\frac{k}{2a_0^2}+\frac{1}{2}}
		\nonumber\\
		&\quad\times\,\int\,dT\,\Bigg[\int_{t(T)}\,dt'\,\dot{\phi}^2(t')\,t'^{-\frac{k}{a_0^2}}\Bigg]\,T^{-\frac{1}{2}-\frac{k}{2a_0^2}} .
	\end{align}
	Equation \eqref{558} yields new analytical $F(T)$ solutions for the following cases:
	\begin{enumerate}
		\item \textbf{General Power law} $\phi(t){\bf =p_0\,t^p}$:
		\begin{align}\label{558a}
			F(T)=&\,-\Lambda_0+B\,T^{\frac{1}{2}-\frac{k}{2a_0^2}}+\frac{\kappa\,p_0^2\,p^2}{2\left(2p-1-\frac{k}{a_0^2}\right)(1-p)}\left(6\left(1- \frac{k}{a_0^2}\right)\right)^{p}\,T^{1-p} .
		\end{align}
		
		\item \textbf{Special Power law} $\phi(t){\bf =p_0\,t^{\frac{1}{2}+\frac{k}{2a_0^2}}}$:
		\begin{align}\label{558b}
			F(T)=&\,-\Lambda_0+B\,T^{\frac{1}{2}-\frac{k}{2a_0^2}}
			\nonumber\\
			&\,+\frac{\kappa\,p_0^2}{8}\,\frac{\left(1+\frac{k}{a_0^2}\right)^2}{\left(1- \frac{k}{a_0^2}\right)^{2}}\left(6\left(1- \frac{k}{a_0^2}\right)\right)^{\frac{k}{2a_0^2}+\frac{1}{2}}\left[2-\left(1-\frac{k}{a_0^2}\right) \ln\! \left(T\right)\right]\,T^{\frac{1}{2}-\frac{k}{2a_0^2}}  .
		\end{align}
		
		\item \textbf{Logarithmic} $\phi(t){\bf =p_0\ln\left(pt\right)}$:
		\begin{align}\label{558c}
			F(T)=&\,-\Lambda_0+B\,T^{\frac{1}{2}-\frac{k}{2a_0^2}}-\frac{\kappa\,p_0^2}{2\left(1+\frac{k}{a_0^2}\right)}\,T .
		\end{align}
		
		\item \textbf{Exponential} $\phi(t){\bf =p_0\exp\left(pt\right)}$:
		\begin{align}\label{558d}
			F(T)=&\,-\Lambda_0+B\,T^{\frac{1}{2}-\frac{k}{2a_0^2}}+\frac{\kappa\,p_0^2\,p^2}{2}\left(6\left(1- \frac{k}{a_0^2}\right)\right)^{\frac{k}{2a_0^2}+\frac{1}{2}}
			\nonumber\\
			&\quad\times\,\int\,dT\,\Bigg[\int_{t(T)}\,dt'\,\exp\left(2pt'\right)\,t'^{-\frac{k}{a_0^2}}\Bigg]\,T^{-\frac{1}{2}-\frac{k}{2a_0^2}} .
		\end{align}
		There is no general solution, but for specific values of $\frac{k}{a_0^2}>0$
		\begin{itemize}
			\item ${\bf \frac{k}{a_0^2}\,\rightarrow\,1}$:
			\begin{align}\label{558da}
				F(T)\approx &\,-\Lambda_0-3\kappa\,p_0^2\,p^2\left(\gamma+\frac{3}{2}\ln(24)\right)\,\epsilon\,\ln(T) ,
			\end{align}
			where $F(T)\,\rightarrow\,-\Lambda_0$ under $\epsilon\,\rightarrow\,0$ limit and {\color{black} a GR solution}.
			
			\item ${\bf \frac{k}{a_0^2}=2}$:
			\small
			\begin{align}\label{558db}
				F(T)=&\,-\Lambda_0+B\,T^{-\frac{1}{2}}-6\kappa\,p_0^2\,p^2 \Bigg[2\sqrt{6} p\,(-T)^{-1/2} {Ei}_{1}\! \left(-2 p \sqrt{6}\,(-T)^{-1/2}\right)
				\nonumber\\
				&\,-\mathrm{Ei}_{1}\! \left(-2 p \sqrt{6}\,(-T)^{-1/2}\right)+{e}^{2 p \sqrt{6}\,(-T)^{-1/2}}-1\Bigg] .
			\end{align}
			\normalsize
			
		\end{itemize}
		There are several possible new $F(T)$ solutions from the Equation \eqref{558} integral with other scalar field sources.		
		
	\end{enumerate}

	\item ${\bf n=2}$: Equation \eqref{554} becomes
	\begin{align}\label{559}
		0=& \frac{a_0^2T}{6k}-\frac{4a_0^2}{k}\,t^{-2}+ t^{-4} ,
		\nonumber\\
		\Rightarrow &\quad t^{-2}(T)= \frac{2a_0^2}{k}\left[1 +\delta_2\sqrt{1-\frac{k}{24a_0^2}\,T}\right] ,
	\end{align}
	where $\delta_2=\pm 1$. Equation \eqref{553} becomes
	\small
	\begin{align}\label{560}
		F(T)=&\,-\Lambda_0+\int\,dT\,\Bigg[C_1-\frac{\kappa}{4}\,\int_{t(T)}\,dt'\,\dot{\phi}^2(t')\,\exp\left(\frac{k\,t'^{-2}}{4\,a_0^2}\right)\Bigg]\left[\frac{2a_0^2}{k}\left[1 +\delta_2\sqrt{1-\frac{k}{24a_0^2}\,T}\right]\right]^{-1/2}
		\nonumber\\
		&\,\times\,\exp\left(-\frac{1}{2}\left[1 +\delta_2\sqrt{1-\frac{k}{24a_0^2}\,T}\right]\right) .
	\end{align}
	\normalsize
	Equation \eqref{560}, for the power-law scalar field $\phi(t)=p_0\,t^p$ and $\frac{k}{a_0^2}=1$, yields new analytical $F(T)$ solutions in the following subcases:
	\begin{itemize}
		\item ${\bf p=-\frac{1}{2}}$:
		\small
		\begin{align}\label{560a}
			F(T)=&\,-\Lambda_0+ \sqrt{1+ \delta_{2} \sqrt{1-\frac{T}{24}}}
			\nonumber\\
			&\quad\times \,\Bigg[C_1\, \exp\left(-\frac{1}{2}\left[1 +\delta_2\sqrt{1-\frac{T}{24}}\right]\right)+\frac{4\kappa\,p_0^2}{\sqrt{2}}\,\left(\delta_{2} \sqrt{1-\frac{T}{24}}-2\right)\Bigg] .
		\end{align}
		\normalsize
		
		\item ${\bf p=-\frac{3}{2}}$:
		\small
		\begin{align}\label{560b}
			F(T)=&\,-\Lambda_0+\sqrt{1+ \delta_{2} \sqrt{1-\frac{T}{24}}}\,\Bigg[C_1\,\exp\left(-\frac{1}{2}\left[1 +\delta_2\sqrt{1-\frac{T}{24}}\right]\right) +\frac{144\kappa\,p_0^2}{\sqrt{2}}
			\nonumber\\
			&\,\times\,\Bigg[\left(\delta_{2} \sqrt{1-\frac{T}{24}}-2\right) +\frac{1}{10}\left(1+\delta_{2} \sqrt{1-\frac{T}{24}}\right)\left(3\delta_{2} \sqrt{1-\frac{T}{24}}-2\right)\Bigg]\Bigg] .
		\end{align}
		\normalsize
		
	\end{itemize}
	There are several other possible $F(T)$ solutions that can be achieved by setting other values of $\frac{k}{a_0}$, $p$ and/or other scalar field expressions. However, we can expect that such cases will yield a more significant expression of $F(T)$.
	
	\item ${\bf n\gg 1}$: Equation \eqref{554} {\color{black} leads to $t(T)= n\,\sqrt{\frac{6}{T}}$ with $\delta_2=1$. From this point, we obtain the exact Equation \eqref{402} formula and then Equations \eqref{403}--\eqref{406} as $F(T)$ solutions for the large values of $n$. The $n\gg 1$ solutions are the same as for the $k=0$ and $-1$ cases and the graphs are shown in Figure \ref{figure1}d.}  We can conclude that the ${\bf k=\pm 1}$ teleparallel $F(T)$ solutions for ${\bf n\rightarrow\,\infty}$ will lead to the flat cosmological (${\bf k=0}$) solutions as a general limit.

\end{enumerate}

All the previous teleparallel $F(T)$ solutions found in Sections \ref{sect3}--\ref{sect42} are new and go further than some recent research papers in the literature. Therefore, there are several other possible subcases which might lead to additional new $F(T)$ solutions and possibly develop in the same direction.%Please check the intended meaning is retained. AL: It is OK.

{\color{black}
	We plotted the $n=1$ and $2$ expanding universe case $F(T)$ solutions, as shown in Figure \ref{figure3}, to compare with the results in Figures \ref{figure1} and \ref{figure2}. The $n=1$ case on subfigure (a) is a similar case to those obtained and presented in Figures \ref{figure1} and \ref{figure2}. The $n=2$ case compares and shows that two power law source cases can be different; this is the case for $p=-\frac{1}{2}$ and $-\frac{3}{2}$ power law sources, as in the similar case shown in Figure \ref{figure2}c. The $n\gg1 $ case is described by Figures \ref{figure1}d and \ref{figure2}d and it is useful for case comparisons.
	
	As for Sections \ref{sect3} and \ref{sect4}, we can still suggest comparing between Figure \ref{figure3} curves and observational data, as carried out in refs. \cite{dixit,ftbcosmo3} by using some usual techniques such as the tensor-to-scalar ratio or others. We can also extend some future works to a thermodynamic-based parameter evolution study, as suggested in ref. \cite{FTBcosmogholamilandry}.
	
	\vspace{-15pt}
	\begin{figure} %[H]
		\hspace{-13pt}\includegraphics[scale=0.80]{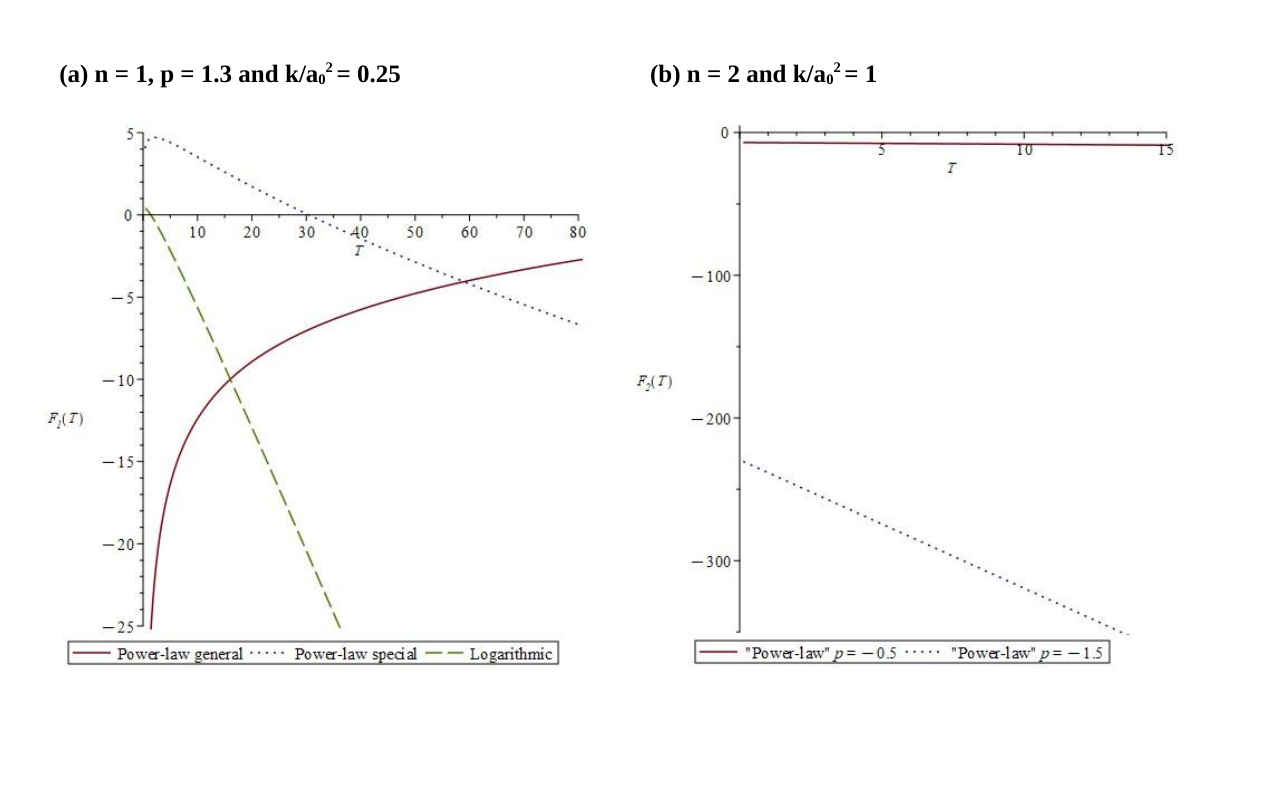}
		\vspace{-43pt}
		\caption{{\color{black}{Plots} %MDPI: Please change the hyphen (-) into a minus sign ($-$, "U+2212"). e.g., "-1" should be "$-$1". AL: Everything is rectified and OK.
				of $F_2(T)=F(T)+\Lambda_0$ versus torsion scalar $T$ described by the $F(T)$ from Equations \eqref{558a}--\eqref{560b} for different types of scalar field sources and $p$ parameters with $\frac{\kappa\,p_0^2}{2}=1$ in the $n=1$ and $n=2$ cases.}}
		\label{figure3}
	\end{figure}
	
}

%\vspace{-10pt}

\newpage

\section{Concluding Remarks}\label{sect5}

The present work first allowed us to solve the TRW $F(T)$ FEs obtained in refs. \cite{preprint,coleylandrygholami} for a scalar field source $\phi$ expressed in terms of density $\rho_{\phi}$ and pressure $P_{\phi}$. For flat cosmological spacetimes (the $k=0$ case), by using the general equation, Equation \eqref{402}, with different scalar fields $\phi(t)$, we obtained purely analytical and simple teleparallel $F(T)$ solutions, as expressed in Equations \eqref{403}--\eqref{406}. This was all made possible by using the power ansatz for the scale factor $a(t)$ and then the relation $t(T)$ obtained with Equation \eqref{400c} to express everything in terms of scalar torsion $T$. One could still use Equation \eqref{402} to obtain other $F(T)$ solutions from other scalar fields $\phi(t)$. This type of result is a logical continuation of the literature, because there are many $k=0$ solutions for other types of sources, not only in teleparallel $F(T)$ gravity, but also in its extensions. The most important thing here is obtaining the all-purpose equation, Equation \eqref{402}, to generate all the teleparallel $F(T)$ solutions needed to study and compare the different $k=0$ cosmological solutions. This flexibility will allow for the study of cosmological phenomena with DE for any values of $n$ and potential $V(\phi)$ involved in the models. These are good advantages favoring our new $F(T)$ solutions.

%\newpage

We have also found general and multipurpose formulas for non-flat cosmological spacetimes (the $k=-1$ and $+1$ cases) from the $k=\pm 1$ TRW $F(T)$ FEs and the power ansatz for $a(t)$, namely Equations \eqref{502} and \eqref{552}, respectively. However, the characteristic {Equations \eqref{504} and \eqref{554}} from the scalar torsion Equations \eqref{500c} and \eqref{550c} ultimately allow only a limited number of purely analytic teleparallel $F(T)$ solutions defined for very specific values of $n$. Simple analytical solutions are obtained for $n=1$ and in the limit of very large $n$, which allows us, in the latter case, to treat models involving the very rapid acceleration of the expansion of the universe. This type of situation will be very useful for the study of phantom energy models and, more precisely, the physical mechanisms leading to the Big Rip. The large $n$ limit solutions for any $k=0$ and $\pm 1$ cases lead to the same {\color{black} {Equation \eqref{403}--\eqref{406}} $F(T)$ solutions for each scalar field source.} For lower values of $n$, the latter will be especially useful for the study of the DE quintessence for non-flat cosmological systems. Note that the cases of early ($n=\frac{1}{2}$) and late ($n=2$) expansion lead to very different teleparallel $F(T)$ solutions: this represents an important benefit for studies of the different forms of DE quintessence. However, the expressions for $F(T)$ often involve various special functions, which is a sign that these are mathematically more sophisticated cases. The $n=1$ solutions constitute an intermediate case between early and late universe expansions that also involves the DE quintessence process to thus better bridge the gap between the $n=\frac{1}{2}$ and $2$ solutions. In principle, other values of $n$ could have been used, but this would only have complicated the present approach without necessarily providing better results and conclusions.

The results of the present work, as well as those obtained in ref. \cite{coleylandrygholami} for linear perfect fluids, will together allow for the full development of TRW $F(T)$-type cosmological models involving different forms of DE. There have been attempts at a similar development to this one in recent papers. These studies often focused on teleparallel extensions \cite{FTBcosmogholamilandry,Kofinas,BohmerJensko,aldrovandi2003,bounce,Capozz}. There have been very recent similar studies involving KS-type spacetimes with, however, fewer symmetries and possible simplifications at the origin ($four$ KVs instead of $six$) \cite{nonvacKSpaper,scalarfieldKS,roberthudsonSSpaper}. The present paper on TRW $F(T)$ solutions {\color{black} is complementary to} the various studies providing the essential ingredients for perfect fluid and scalar field {\color{black} teleparallel} $F(T)$ solution classes. {\color{black} As in recent works, the primary aims and scopes of this study are purely theoretical~{\cite{nonvacKSpaper,scalarfieldKS,FTBcosmogholamilandry}.} As suggested in Sections \ref{sect3}--\ref{sect42} and in ref. \cite{FTBcosmogholamilandry}, the new $F(T)$ should be compared with observational data by some fitting techniques, as conducted in refs. \cite{dixit,ftbcosmo3}. By this method, we will be able to determine the exact values of the parameters involved in the cosmological models.} We must keep in mind that the ultimate aim is the complete study of teleparallel cosmological models involving the various DE forms. These are therefore the next steps in the development of teleparallel gravity and we must now go all the way.

%\newpage

\section*{Abbreviations}

\noindent The following abbreviations are used in this manuscript:\\
\begin{tabular}{@{}llll}
AL & Alexandre Landry & CK & Cartan--Karlhede  		\\  
DE & Dark Energy  	& EoS & Equation of State 	\\ 
%MDPI: Please confirm if it should be removed. AL: It is removed.
FE & Field Equation 	& GR & General Relativity 	\\  
KS & Kantowski--Sachs & KV & Killing Vector \\
NGR & New General Relativity & RW &  Robertson--Walker\\
TEGR & Teleparallel Equivalent of General Relativity & {\color{black} TdS} & {\color{black} Teleparallel de Sitter}\\ 
TRW & Teleparallel Robertson--Walker & & \\ 
\end{tabular}

\end{document}